\documentclass[%
 preprint,epsfig,showkeys,
bibnotes,
 amsmath,amssymb,
 aps, physrev,
floatfix,
]{revtex4-2}

\usepackage{graphicx}
\usepackage{dcolumn}
\usepackage{bm}

\begin{document}

\title{\textbf{New
 Perspectives on the Irregular Singular Point of the Wave Equation for a Massive Scalar Field in
Schwarzschild Space-Time} 
}

\author{Giampiero Esposito}
\affiliation{Dipartimento di Fisica ``Ettore Pancini'', Complesso Universitario di Monte S. Angelo,
Via Cintia Edificio 6, 80126 Napoli, Italy \\
\quad Istituto Nazionale di Fisica Nucleare, Sezione di Napoli, Complesso Universitario di Monte S. Angelo,
Via~Cintia Edificio 6, 80126 Napoli, Italy}

\affiliation{giampiero.esposito@na.infn.it\\ORCID: 0000-0001-5930-8366}
\author{Marco Refuto}%
\affiliation{%
 Dipartimento di Fisica ``Ettore Pancini'', Complesso Universitario di Monte S. Angelo,
Via Cintia Edificio 6, 80126 Napoli, Italy}
 \affiliation{unimarcoref@gmail.com \\ ORCID: 0009-0007-6563-7909}

\begin{abstract}
For a massive scalar field in a fixed Schwarzschild background, the 
radial wave equation obeyed by Fourier modes is first studied. 
After reducing such a radial wave equation to its normal form, we first
study approximate solutions in the neighborhood of the origin, horizon and point
at infinity, and then we relate the radial with the Heun equation, obtaining 
local solutions at the regular singular points.
Moreover, we obtain the full asymptotic expansion of the local solution 
in the neighborhood of the irregular singular point at infinity. We also 
obtain and study the associated integral representation of the massive scalar field. 
Eventually, the technique developed for the irregular singular point is applied 
to the homogeneous equation associated with the inhomogeneous Zerilli equation 
for gravitational perturbations in a Schwarzschild background.
\end{abstract}

\maketitle

\tableofcontents

\section{{Introduction}
}

Almost a hundred and ten years after the creation of general relativity~\cite{Einstein1,Einstein2},
even the study of just the scalar wave equation in a fixed curved background,
which solves the Einstein equations, is still in the early stages. In~the
seventies, for~example, very detailed papers were published on the radial
wave equation in Schwarzschild space-time, but~they claimed that it is not related 
to any known differential equation of mathematical physics~\cite{Persides1,Persides2,Persides3}. Forty years after this work,
the dissertation in Ref.~\cite{Wardell} was still claiming that
no exact solution of the radial wave equation is known in Schwarzschild
space-time.

Every partial differential equation studied in physics and mathematics
is a world of its own, and~the choice of the appropriate techniques depends
on the assumptions made: the nature of the physical phenomenon, its
mathematical formulation, the~ambient space-time manifold, the~regularity
required for the solutions and the existence or lack of symmetries.
It has been, therefore, our aim to exploit the well-established tools of
ordinary and partial differential equations to obtain both the exact solution 
of the radial wave equation for Fourier modes of a massive scalar field
and the integral representation of the full scalar field in Schwarzschild
geometry. The~scalar field case is already of considerable physical interest  
in~light of modern investigations of null infinity~\cite{Schmidt,Kehrberger} 
and of its role in studying Hawking radiation~\cite{Hawking} for
quantum fields in curved space-time~\cite{Figari,Witten}. The~spherical 
symmetry of the fixed space-time geometry makes it possible to exploit 
Fourier modes, which do not exist in a generic space-time without any~symmetry.

The paper is organized as follows. Section~\ref{222} is devoted to an original
synthesis of the radial wave equation and its reduction to normal form.
Section~\ref{333} obtains approximate solutions of such an equation in the
neighborhood of two regular singular points and of 
an irregular singular point. Section~\ref{444} gives the radial wave
equation in the Heun equation form. Local solutions at the regular singular
points are obtained in Section~\ref{555}. A~full local solution 
of the radial wave equation at infinity is obtained in Section~\ref{666}.
The resulting integral representation 
of the massive scalar field that solves the wave equation is studied
analytically and numerically in Section~\ref{777}. Section~\ref{888} is devoted
to the application of Section~\ref{666} to the investigation of the homogeneous
equation associated with the inhomogeneous Zerilli equation for
gravitational perturbations of a Schwarzschild background.
The results obtained and open problems
are presented in Section~\ref{999}, and~relevant details are 
provided in the~appendices.

\section{Radial Wave Equation and Its Reduction to Normal~Form}\label{222}

In agreement with our aims, our starting point is the scalar wave {equation}
\begin{equation}
(\Box -\mu^{2})\psi=0
\label{(2.1)}
\end{equation}
for a scalar field $\psi$ of mass $\mu$ in a fixed Schwarzschild background
(thus, we are not studying a coupled system of Einstein and scalar-field
equations, nor are we studying a scalar self-force problem). 
This background space-time is the unique spherically
symmetric ({{This} 
 means that space-time admits the group 
$SO(3)$ as a group of isometries, with~the group orbits given by
spacelike two-surfaces.})
solution of the vacuum Einstein equations in a way made
precise by the Birkhoff theorem~\cite{HE}. In~standard spherical
coordinates $(t,r,\theta,\phi)$ where 
$t \in {\bf R} \cup \{-\infty \} \cup \{+\infty \}, r \geq 0,
\theta \in [0,\pi], \phi \in [0,2\pi]$, 
with $G=c=1$ units, on~defining
$f(r)=1-\frac{2M}{r}$, the~space-time metric is diagonal:
\begin{equation}
g=-f(r) dt^{2}+\frac{dr^{2}}{f(r)}+r^{2}(d\theta^{2}+\sin^{2}\theta d\phi^{2}).
\label{(2.2)}
\end{equation}
{Since}
 such components are $t$-independent, one can exploit a Fourier
representation according to the expansion~\cite{Wardell}
($Y_{lm}$ being the spherical harmonics on the $2$-sphere, and~$C_{lm}$ being coefficients of linear 
combination {(}{The $C_{lm}$ coefficients can be derived by inverse
Fourier transform following standard techniques, but~we do not need their
values in our analysis}{)}).
\begin{equation}
\psi(t,r,\theta,\phi;\mu)=\sum_{l=0}^{\infty}\psi_{l \mu}(t,r)\sum_{m=-l}^{l}
C_{lm}Y_{lm}(\theta,\phi), 
\label{(2.3)}
\end{equation}
where $\psi_{l \mu}$ are the integrated Fourier modes
\begin{equation}
\psi_{l \mu}(t,r)=\int_{-\infty}^{\infty}e^{-i \omega t}
R_{l \omega \mu}(r) \frac{d\omega}{\sqrt{2\pi}},
\label{(2.4)}
\end{equation}
and $R_{l\omega \mu}$ solves the radial wave equation
\begin{equation}
\left[\frac{d^{2}}{dr^{2}}+\frac{2(r-M)}{r^{2}f(r)}\frac{d}{dr}
+\frac{1}{f(r)}\left(\frac{\omega^{2}}{f(r)}-\frac{l(l+1)}{r^{2}}
-\mu^{2}\right)\right]R_{l \omega \mu}(r)=0.
\label{(2.5)}
\end{equation}
{This} equation can be written in normal form by assuming the product
\begin{equation}
R_{l\omega \mu}(r)=\alpha(r)\beta_{l\omega \mu}(r),
\label{(2.6)}
\end{equation}
and looking for an $\alpha(r)$ such that the equation obeyed by
$\beta_{l\omega \mu}(r)$ has vanishing coefficient of the first derivative.
Indeed, upon~setting in Equation \eqref{(2.5)}
\begin{equation}
p(r)=\frac{2(r-M)}{r^{2}f(r)}, \;
q_{l\omega \mu}(r)=\frac{1}{f(r)}\left(
\frac{\omega^{2}}{f(r)}-\frac{l(l+1)}{r^{2}}
-\mu^{2}\right),
\label{(2.7)}
\end{equation}
the general method for achieving normal form~\cite{Esposito} yields
\begin{equation}
\alpha(r)={\rm exp} \left(-\frac{1}{2}\int p(r)dr \right)
=\frac{1}{\sqrt{r(r-2M)}},
\label{(2.8)}
\end{equation}
jointly with the second-order differential equation
\begin{equation}
\left[\frac{d^{2}}{dr^{2}}+J_{l\omega \mu}(r)\right]\beta_{l\omega \mu}(r)=0,
\label{(2.9)}
\end{equation}
where the potential term is given by
\begin{eqnarray}
\; & \; & J_{l\omega \mu}(r)=-\frac{1}{2}\frac{dp}{dr}
-\frac{1}{4}p^{2}(r)+q_{l\omega \mu}(r)
\nonumber \\
&=& \frac{1}{4r^{2}}+\frac{1}{4(r-2M)^{2}}
-\frac{\left(\frac{1}{2}+l(l+1)\right)}{r(r-2M)}
\nonumber \\
&+& \frac{\omega^{2}r^{2}}{(r-2M)^{2}}
-\frac{\mu^{2}r}{(r-2M)}.
\label{(2.10)}
\end{eqnarray}
{We} note that $J_{l\omega \mu}(r)$ has a second-order pole at $r=0$ and at
$r=2M$, and~hence these are regular singular points, whereas
the point $r=\infty$ is an irregular singular point~\cite{Forsyth} 
as we are going to prove. In~order to study the nature of the $r=\infty$
point, one defines the new independent variable $\rho=\frac{1}{r}$, which
implies for $\beta_{l\omega \mu}(\rho)$ the differential equation
\begin{equation}
\left[\frac{d^{2}}{d\rho^{2}}+\frac{2}{\rho}
+\frac{J_{l\omega \mu}(\rho)}{\rho^{4}}\right]\beta_{l\omega \mu}(\rho)=0,
\label{(2.11)}
\end{equation}
where
\begin{equation}
\frac{J_{l\omega \mu}(\rho)}{\rho^{4}}=\gamma_{1l \omega}(\rho)
+\gamma_{2\omega \mu}(\rho),
\label{(2.12)}
\end{equation}
having defined
\begin{equation}
\gamma_{1l \omega}(\rho)=\frac{1}{4\rho^{2}}\left(1
+\frac{1}{(1-2M \omega)^{2}}\right)
-\frac{\left(\frac{1}{2}+l(l+1)\right)}{\rho^{2}(1-2M\rho)},
\label{(2.13)}
\end{equation}
\begin{equation}
\gamma_{2\omega \mu}(\rho)=\frac{1}{\rho^{4}}
\left(\frac{\omega^{2}}{(1-2M \rho)^{2}}
-\frac{\mu^{2}}{(1-2M \rho)}\right).
\label{(2.14)}
\end{equation}
{Since} $\gamma_{2\omega \mu}(\rho)$ has a fourth-order pole at $\rho=0$, 
the~point $r=\infty$ is indeed an irregular singular point for Equation \eqref{(2.9)}.

The material in this section is standard but~nevertheless useful for achieving 
a self-contained presentation of our subsequent~analysis.

\section{Approximate Solutions in the Neighborhood of \boldmath{$r=0,2M$} and \boldmath{$\infty$}}\label{333}

\subsection{Fourier Modes as $r \rightarrow 0$}

As $r \rightarrow 0$, the~form of $J_{l\omega \mu}(r)$ in Equation \eqref{(2.10)}
implies that Equation \eqref{(2.9)} reduces to a form independent of $l$ and
$\omega$, i.e.,

\begin{equation}
\left[\frac{d^{2}}{dr^{2}}+\frac{1}{4r^{2}}\right]\beta_{0}(r)=0.
\label{(3.1)}
\end{equation}
{On} looking for solutions in the form $\beta_{0}(r)=r^{\alpha}$, one finds
that the square of $\left(\alpha-\frac{1}{2}\right)$ should vanish. 
Hence there exist two linearly independent integrals giving rise to
\begin{equation}
\beta_{0}(r)=C_{1}\sqrt{r}+C_{2}\sqrt{r}\log(r).
\label{(3.2)}
\end{equation}
{On} requiring regularity at the origin one finds $C_{2}=0$, and~hence
\begin{equation}
R_{0}(r)=\frac{\beta_{0}(r)}{\sqrt{r(r-2M)}}=\frac{C_{1}}{\sqrt{r-2M}},
\label{(3.3)}
\end{equation}
i.e., the~Fourier modes are finite and constant at $r=0$.
This agrees with the property in the massless case found in Ref.
~\cite{Persides1}: there exists one solution finite at the origin,
which is physically preferable for this~reason.

\subsection{Fourier Modes as $r \rightarrow 2M$}

As $r \rightarrow 2M$, by~virtue of Equation \eqref{(2.10)} we find that
\begin{equation}
\lim_{r \to 2M}J_{l\omega \mu}(r)=\lim_{r \to 2M}\left[
\frac{\chi_{2}(\omega)}{(r-2M)^{2}}
+\frac{\chi_{1}(l,\mu)}{(r-2M)}
+\chi_{0} \right],
\label{(3.4)}
\end{equation}
where
\begin{equation}
\chi_{2}(\omega)=4M^{2}\omega^{2}+\frac{1}{4},
\label{(3.5)}
\end{equation}
\begin{equation}
\chi_{1}(l,\mu)=-\frac{1}{2M}\left(\frac{1}{2}+l(l+1)\right)-2M \mu^{2},
\label{(3.6)}
\end{equation}
\begin{equation}
\chi_{0}=\frac{1}{16M^{2}}.
\label{(3.7)}
\end{equation}
{This} implies that Equation \eqref{(2.9)} reduces to
\begin{equation}
\left[\frac{d^{2}}{dr^{2}}
+\frac{\chi_{2}(\omega)}{(r-2M)^{2}}\right]\beta_{\omega,2M}(r)=0,
\label{(3.8)}
\end{equation}
which (up to a multiplicative constant) is solved by
\begin{equation}
\beta_{\omega,2M}(r)=(r-2M)^{A},
\label{(3.9)}
\end{equation}
where $A$ is a root of the algebraic equation
\begin{equation}
A^{2}-A-\chi_{2}(\omega)=0.
\label{(3.10)}
\end{equation}
{Such} roots are
\begin{equation}
A_{+}=\frac{1}{2}+2iM\omega,
\label{(3.11)}
\end{equation}
\begin{equation}
A_{-}=\frac{1}{2}-2iM\omega,
\label{(3.12)}
\end{equation}
and hence the Fourier modes have the limiting form
\begin{eqnarray}
\; & \; & R_{\omega,2M}(r)=\frac{\beta_{\omega,2M}(r)}{\sqrt{r(r-2M)}}
\nonumber \\
&=& \frac{1}{\sqrt{r}}\Bigr[D_{1}(r-2M)^{2iM \omega}
+D_{2}(r-2M)^{-2iM \omega}\Bigr],
\label{(3.13)}
\end{eqnarray}
where $D_1$ and $D_2$ are suitable constants. Interestingly, this
limiting behavior agrees with the massless case~\cite{Persides1}:
the two solutions, with~coefficients $D_1$ and $D_2$, remain
bounded on the horizon $r=2M$, and~no solution approaches zero 
as $r \rightarrow 2M$. These properties are closely related
to the possibility of destroying the black hole and to the
radiation of higher multipole moments when a small scalar particle
falls into the black hole~\cite{Cohen}.

\subsection{Fourier Modes at~Infinity}

Last but not least, the~potential term $J_{l \omega}$ in Equation
\eqref{(2.9)} has the following Poincar\'e asymptotic expansion 
as $r \rightarrow \infty$:
\begin{equation}
J_{l \omega \mu} \sim \omega^{2}-\mu^{2}
+\frac{A_{1}(\omega,\mu)}{r}
+\frac{A_{2}(l,\omega,\mu)}{r^{2}}
+\sum_{k=3}^{\infty}\frac{A_{k}(l,\omega,\mu)}{r^{k}}
\label{(3.14)}
\end{equation}
where we have defined
\begin{equation}
A_{1}(\omega,\mu)=2M(2 \omega^{2}-\mu^{2}),
\label{(3.15)}
\end{equation}
\begin{equation}
A_{2}(l,\omega,\mu)=4M^{2}(3 \omega^{2}-\mu^{2})-l(l+1).
\label{(3.16)}
\end{equation}
Thus, as~$r \rightarrow \infty$, 
$\beta_{l \omega \mu}(r)$ is (up to a multiplicative constant) 
the product of the leading order
$\beta_{\omega,\infty}(r)=e^{\pm r \sqrt{\mu^{2}-\omega^{2}}}$
times a function that we do not strictly need here, but~is part
of an analysis that we will perform in Section~\ref{666}.

Note also that, by~virtue of Equation \eqref{(2.4)}, 
one has to integrate overall values of $\omega$ in order to evaluate the
large-$r$ scalar field. Thus, the~Fourier modes
at large $r$ have to be considered
in the following intervals for the $\omega$ variable:
$$
[-\infty,-\mu], \; ]-\mu,\mu[ \; {\rm and} \; [\mu,\infty].
$$
This remark will be exploited in Section~\ref{777}.

\section{Connecting the Radial with the Heun~Equation}\label{444}

The reduction to canonical form of Equation \eqref{(2.5)} helped us in
searching for the singular points $r_{0}=0$, $r_{2M}=2M$ and for
the one at infinity. Now we are going to see that Equation~\eqref{(2.5)}
belongs to the class of confluent Heun 
equations defined in Appendix \ref{appA}~\cite{Ronveaux}. In~order to prove
it, we rely upon the change of independent variable frequently used
in the literature, i.e.,
\begin{equation}
z=\frac{(r-r_{2M})}{(r_{0}-r_{2M})}
=1-\frac{r}{2M}.
\label{(4.1)}
\end{equation}
{In} this way, the~two finite singularities $r=\{0,2M \}$ correspond to
$z=\{1,0 \}$, respectively, and~the physical requirement $r \geq 0$
leads to the restriction $z \leq 1$. The~Equation \eqref{(2.5)}
henceforth takes the form
\begin{equation}
\left[ \frac{d^{2}}{dz^{2}}
-\frac{(1-2z)}{z(1-z)} \frac{d}{dz} 
+ \frac{4M^{2}\omega^{2}(1-z)^{2}}{z^{2}}
+\frac{l(l+1)}{z(1-z)}
+\frac{4M^{2}\mu^{2}(1-z)}{z}\right]R_{l\omega \mu}(z)=0.
\label{(4.2)}
\end{equation}
{Thanks} to this change of variable, such equation is nothing but a
confluent Heun equation in the so-called non-symmetric canonical 
form. Such a form is obtained by exploiting
\begin{equation}
H_{{\rm NSCF}}(z)=e^{\alpha_{1}z} \; z^{\alpha_{2}} 
\; (z-1)^{\alpha_{3}} \; H(z),
\label{(4.3)}
\end{equation}
where $H(z)$ solves the confluent Heun~equation.

\section{Local Solutions at the Regular Singular~Points}\label{555}

In our specific case, upon~defining the five Heun parameters
(no confusion should arise with $\alpha$ and $\beta$ of {Section}~\ref{222})
\begin{eqnarray}
\; & \; & \alpha=4M \sqrt{\mu^{2}-\omega^{2}}, 
\nonumber \\
& \; & \beta=4i M \omega,
\nonumber \\
& \; & \gamma=0,
\nonumber \\
& \; & \delta=4M^{2}(\mu^{2}-2 \omega^{2}),
\nonumber \\
& \; & \eta=-4M^{2}(\mu^{2}-2 \omega^{2})-l(l+1),
\label{(5.1)}
\end{eqnarray}
{Equation} \eqref{(4.3)} takes the form
\begin{equation}
H_{{\rm NSCF}}(z)=e^{\frac{\alpha}{2}z} \;
z^{\frac {\beta}{2}} H(z),
\label{(5.2)}
\end{equation}
since $\alpha_{3}=0$ for Equation \eqref{(4.2)}. Starting from Equation \eqref{(5.2)}, one can construct the complete Frobenius solution
at $z=0$ (corresponding to $r=2M$), that reads
(cf. Refs.~\cite{Fiziev,Philipp,Vieira,Chen,Minucci})
\begin{equation}
R_{0}(z)= e^{\frac {\alpha}{2}z} \; z^{\frac{\beta}{2}}
\Bigr[C_{1}{\rm HeunC}(\alpha,\beta,\gamma,\delta,\eta,z)
+ C_{2}z^{-\beta}{\rm HeunC}(\alpha,-\beta,\gamma,\delta,\eta,z)\Bigr],
\label{(5.3)}
\end{equation}
whose leading order is
\begin{equation}
R_{0}(z) \sim e^{\frac {\alpha}{2}z} \;
z^{\frac {\beta}{2}} \Bigr(C_{1}+C_{2}z^{-\beta}\Bigr).
\label{(5.4)}
\end{equation}
{The} Frobenius solution at $z=1$ (i.e., at~$r=0$) is given by
\begin{eqnarray}
R_{1}(z)&=& e^{\frac {\alpha}{2}z} \; z^{\frac{\beta}{2}}
\Bigr[C_{1}{\rm HeunC}(-\alpha,\gamma,\beta,-\delta,\eta+\delta,-z)
\nonumber \\
&+& C_{2}(z-1)^{-\gamma}{\rm HeunC}(-\alpha,-\gamma,\beta,-\delta,\eta + \delta,z)\Bigr],
\label{(5.5)}
\end{eqnarray}
whose leading order is
\begin{equation}
R_{1}(z) \sim e^{\frac {\alpha}{2}z} \;
z^{\frac {\beta}{2}} \Bigr(C_{1}+C_{2}(z-1)^{-\gamma}\Bigr).
\label{(5.6)}
\end{equation}

\begin{figure}
\centering
\includegraphics[width=0.49\linewidth]{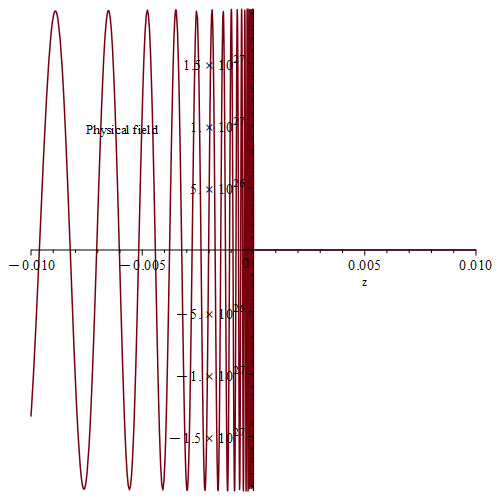}
\caption{{In this plot}, the~real part of Equation \eqref{(5.3)} is shown for 
$z \in [-0.01,0.01]$. The~$s$-wave near $z = 0$ (i.e., $r=2M$) 
is very close to zero inside the black 
hole and diverges on the event horizon,  
and it exhibits strong oscillations.}
\label{fig1}
\end{figure}

\begin{figure}
\centering
\includegraphics[width=0.5\linewidth]{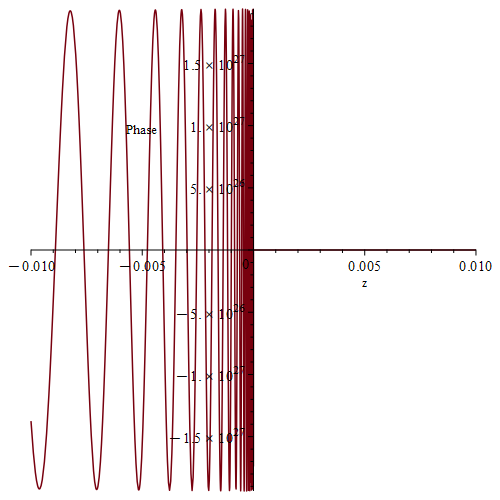}
\caption{{The}  imaginary part of Equation \eqref{(5.3)} is shown for 
$z \in [-0.01,0.01]$ . The~phase of the $s$-wave near $z=0$ 
behaves qualitatively in the same manner as the real part.}
\label{fig2}
\end{figure}

\begin{figure}
\centering
\includegraphics[width=0.5\linewidth]{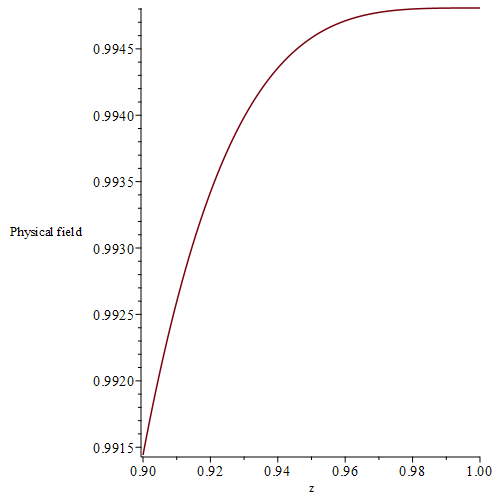}
\caption{The real part of Equation \eqref{(5.5)} is shown for 
$z \in [0.9,1]$. Near~$z = 1$ (i.e., $r = 0$) the 
$s$-wave exhibits a plateau, then an overall downward trend as $z$ 
decreases.}
\label{fig3}
\end{figure}

\begin{figure}
\centering
\includegraphics[width=0.5\linewidth]{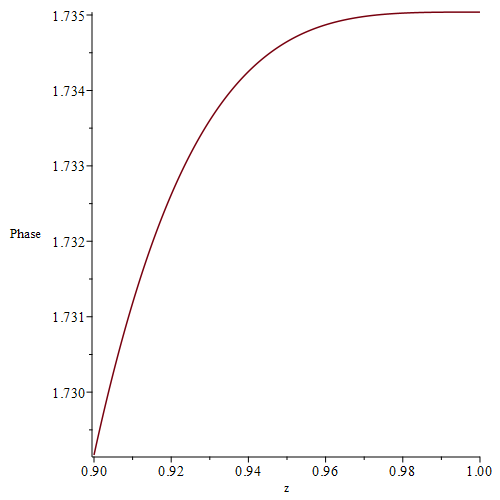}
\caption{The imaginary part of Equation \eqref{(5.5)} is shown for 
$z \in [0.9,1]$. The~phase shows qualitatively the same behavior as 
the real part but with slightly higher values.}
\label{fig4}
\end{figure}

In order to visualize the results obtained so far, we plot in Figures 
\ref{fig1} and \ref{fig2} the real and imaginary part of Equation \eqref{(5.3)},
respectively. Similarly, we plot in Figures \ref{fig3} and \ref{fig4}
the real and imaginary part of Equation \eqref{(5.5)}, respectively.
In all these Figures, we make the following choice for the free parameters:
\begin{equation}
M=10, \quad \mu=0.1, \quad \omega=1, \quad l=0, \quad C_1=C_2=1.
\label{plot parameters}
\end{equation}

In order to summarize the physical content of our results regarding the $s$-wave, 
the real part of the local solution slightly inside the black hole
is very close to zero, while outside the event horizon, it exhibits strong 
oscillations, both in amplitude and in frequency. The~imaginary part of the same solution follows
the same behavior both qualitatively and quantitatively. Moreover, both 
the real and imaginary parts of the local solution at the origin possess a plateau very close to the
z = 1 (i.e., $r=0$) point, and~then decrease as z~decreases.

\unskip
\unskip
\unskip

\section{Full Local Solution for Fourier Modes in the Neighborhood of \boldmath{$r=\infty$}}\label{666}

When $r \rightarrow \infty$ one can evaluate $\beta_{l \omega \mu}(r)$, and~hence the Fourier modes $R_{l \omega \mu}(r)$, by~completing the analysis
initiated in Equations \eqref{(3.14)}--\eqref{(3.16)}. For~this purpose, we
recognize from such equations that the differential Equation 
\eqref{(2.9)} for $\beta_{l \omega \mu}(r)$ can be written in the form
\begin{equation}
\left(\frac{d^{2}}{dr^{2}}+\sum_{k=0}^{\infty}
\frac{A_{k}(l,\omega,\mu)}{r^{k}}\right)\beta_{l \omega \mu}(r)=0,
\label{(6.1)}
\end{equation}
where $A_{0}=\omega^{2}-\mu^{2}$, while $A_{1}$ and $A_{2}$
have been already defined in Equations \eqref{(3.15)}
and \eqref{(3.16)}, respectively.
All subsequent coefficients can be computed from the asymptotic
expansion of Equation \eqref{(2.10)} at large $r$. At~this stage, Equation \eqref{(6.1)} suggests looking for $\beta_{l \omega \mu}(r)$ 
at large $r$ in the form (setting $\varepsilon= \pm 1$ and considering 
$\zeta$ as a parameter to be determined as shown~below)
\begin{equation}
\beta_{l \omega \mu}(r) \sim e^{\varepsilon r \sqrt{-A_{0}}} \; r^\zeta
\; \left(1+\sum_{s=1}^{\infty}
\frac{B_{s}(l,\omega,\mu)}{r^{s}}\right).
\label{(6.2)}
\end{equation}
{The} insertion of Equation \eqref{(6.2)} into Equation \eqref{(6.1)} yields
\begin{equation}
\gamma_{0}+\frac{\gamma_{1}}{r}
+\frac{\gamma_{2}}{r^{2}}+\sum_{k=3}^{\infty}
\frac{\gamma_{k}}{r^{k}}=0.
\label{(6.3)}
\end{equation}
{For} this equation to be identically satisfied as $r \rightarrow \infty$,
all $\gamma_{k}$ coefficients should vanish. Indeed, we find
\begin{equation}
\gamma_{0}=A_{0}-A_{0}=0,
\label{(6.4)}
\end{equation}
while
\begin{equation}
\gamma_{1}=2\zeta \varepsilon \sqrt{-A_0}+A_1,
\label{(6.5)}
\end{equation}
\begin{equation}
\gamma_{2}=(\zeta -1)(2\varepsilon B_1\sqrt{-A_0}+\zeta)+A_1B_1+A_2,
\label{(6.6)}
\end{equation}
jointly with infinitely many other equations for all subsequent
$\gamma_{k}$ coefficients. Since they should all vanish, we obtain
linear algebraic equations for $B_{1},B_{2},...,B_{\infty}$. In~addition, 
since we look for a bounded solution at large $r$, we restrict ourselves 
to the $\varepsilon=-1$ case. {(}{We are here studying a local
solution which only holds at large $r$ \cite{Persides1,Persides3}. 
Thus, the~$\varepsilon=1$ mode
cannot be avoided by studying conditions near the horizon $r=2M$.}{)}
For instance, from Equation~\eqref{(6.5)} we evaluate 
the explicit form of $\zeta$, while from Equation \eqref{(6.6)} we obtain $B_1$,~i.e.,

\begin{equation}
\zeta=\frac{A_1}{2\sqrt{-A_0}}=\frac{M(2\omega^2-\mu^2)}{\sqrt{\mu^2-\omega^2}},
\label{(6.7)}
\end{equation}
\begin{equation}
B_1=\frac{\zeta(\zeta-1)+A_2}{2(\zeta-1)\sqrt{-A_0}-A_1}.
\label{(6.8)}
\end{equation}
{By} virtue of Equation \eqref{(2.6)}, we find eventually the large-$r$
Fourier modes in the form (cf. Refs.~\cite{Fiziev,Philipp,Minucci})
\begin{equation}
R_{l \omega \mu,\infty}(r) \sim \frac{e^{- r \sqrt{-A_{0}}}}
{\sqrt{r(r-2M)}}r^\frac{A_1}{2\sqrt{-A_0}}
\left(1+\sum_{s=1}^{\infty}
\frac{B_{s}(l,\omega,\mu)}{r^{s}}\right).
\label{(6.9)}
\end{equation}
{Note} that, by~considering only the $k=0,1,2$ terms in Equation \eqref{(6.1)}, one would obtain 
the Whittaker equation (a modified form of the confluent hypergeometric equation), solved 
by the two Whittaker functions. However, all negative powers of $r$ occur in Equation \eqref{(6.1)},
and hence one needs our factorization in Equation \eqref{(6.9)}, where the exponential cancels the
effect of $A_{0}$, and~the $B_{s}$ functions of $l,\omega,\mu$ compensate the negative powers
of $r$ in the operator provided that a non-integer power of $r$ is included in our
factorized ansatz. This procedure agrees with the general method of Poincar\'e for linear differential
equations~\cite{Poincare,Poincare2}.

\begin{figure}
\centering
\includegraphics[width=0.52\linewidth]{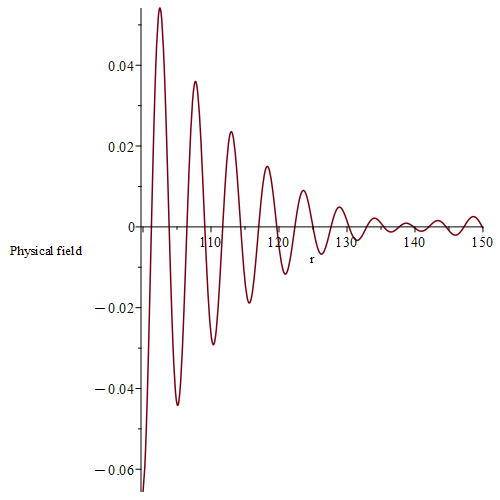}
\caption{{The real part of Equation \eqref{(6.9)} is plotted with parameters given by 
Equation \eqref{plot parameters} summing over $s$ from $1$ 
to $5$ for  $r \in [10M, 15M]$. 
The oscillations of the field become smaller and smaller as $r$ grows.}}
\label{fig5}
\end{figure}

\begin{figure}
\centering
\includegraphics[width=0.5\linewidth]{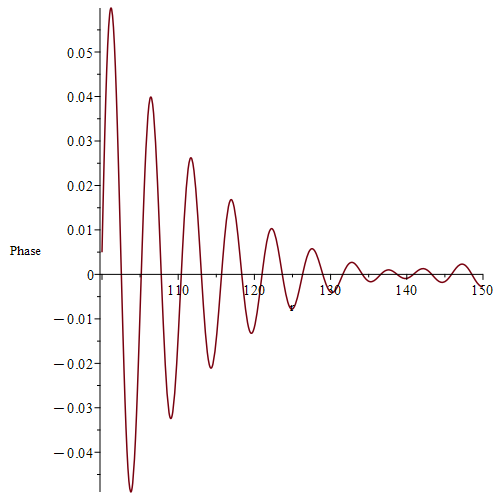}
\caption{The imaginary part of Equation \eqref{(6.9)} is plotted with parameters given by 
Equation \eqref{plot parameters} summing over $s$ from $1$ to $5$ for   
$r \in [10M, 15M]$. The~phase follows qualitatively 
the same behavior of the real part.}
\label{fig6}
\end{figure}

In Appendix \ref{appC} we offer a theoretical motivation for the ansatz given by Equation 
\eqref{(6.2)} and the explicit computation of $\zeta$ and the first 
$A_k$, $B_s$ coefficients. In~order to visualize such a solution, we consider 
Equation \eqref{(6.9)}  with the same choice for parameters given by Equation 
\eqref{plot parameters} (this time we do not have integration constants like 
$C_1$ and $C_2$), ending up with two plots: one for the real part 
(Figure  \ref{fig5}) of the field and the other for the imaginary one (Figure  \ref{fig6}).
The relation with the local results by Persides in the massless case~\cite{Persides1} is discussed in Appendix~\ref{appD}.

\vspace{-2pt}

The plots obtained in our large-$r$ analysis clearly describe the vanishing of the field 
(both real and imaginary parts) at spacelike infinity due to damped 
oscillations. This is in agreement with our mathematical ansatz and physical~requirements.

\section{Large-\boldmath{$r$} Solution of the Full Wave~Equation}\label{777}

So far, we have studied the behavior of solutions for fixed frequencies. 
Since one of our goals is to study the spectrum of a massive field far away 
from the black hole, we now focus the attention on the study of 
the solution at spacelike infinity by including 
the contribution of all frequencies (neglecting the angular part).
In the previous sections we have discovered that when $r \rightarrow \infty$, 
the Fourier modes take the approximate form \eqref{(6.9)}.
The large-$r$ form of the scalar field
$\psi$ which solves Equation \eqref{(2.1)} and satisfies the asymptotic~condition
\begin{equation}
\lim_{r \to \infty} |\psi_{l}(t,r)| < \infty
\label{(7.1)}
\end{equation}
is therefore expressed by Equation \eqref{(2.4)}, where, upon~defining
\begin{equation}
\varphi(t,r,\omega)=\omega t 
+ r \sqrt{\omega^{2}-\mu^{2}}+M\frac{(2\omega^2-\mu^2)}{\sqrt{\omega^2-\mu^2}}\log(r),
\label{(7.2)}
\end{equation}
we find, for~$\omega^2>\mu^2$,
\begin{eqnarray}
\; & \; & \psi^I_{l \mu}(t,r) \sim \int_{-\infty}^{\infty}
e^{-i \omega t}R_{l\omega \mu,\infty}(r)
\frac{d\omega}{\sqrt{2\pi}}
\nonumber \\
& \sim & \frac{1}{\sqrt{r(r-2M)}}
\left(\int_{-\infty}^{-\mu}+\int_{\mu}^{\infty} \right)
\frac{d\omega}{\sqrt{2\pi}}e^{-i \varphi(t,r,\omega)}\left(1+\sum_{s=1}^{\infty}
\frac{B_{s}(l,\omega,\mu)}{r^{s}}\right).
\label{(7.3)}
\end{eqnarray}
{The} integral is qualitatively different in the $\omega^2<\mu^2$ case, since the last two terms 
in Equation \eqref{(7.2)} become imaginary, leading~to
\begin{equation}
\psi^{II}_{l \mu}(t,r) \sim \frac{1}{\sqrt{r(r-2M)}}\int_{-\mu}^\mu
\frac{d\omega}{2\pi}e^{-i\omega t}e^{r\sqrt{\mu^2-\omega^2}
\left(1+M\frac{\mu^2-2\omega^2}{\sqrt{\mu^2-\omega^2}}
\frac{\log(r)}{r}\right)}\left(1+\sum_{s=1}^{\infty}
\frac{B_{s}(l,\omega,\mu)}{r^{s}}\right).
\label{(7.4)}
\end{equation}

\begin{figure}
\centering
\includegraphics[width=0.5\linewidth]{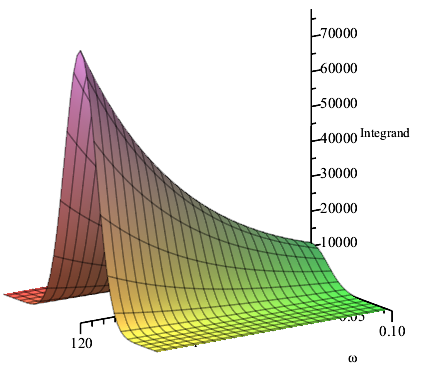}
\caption{{Plot} 
 of Re$\left[\Omega_{0 \mu}^{II}(1,r,\omega)\right]$ for 
$r\in[100,120]$, $\omega \in [-\mu, \mu]$, setting $\mu=0.1$ and $M=10$, 
up to the $r^{-5}$ term. Along the $\omega$-axis the function has a 
Gaussian-like behavior (it is possible to prove it by expanding in series
Equation~\eqref{(7.6)} for small frequencies), while it diverges along the $r$-axis. 
The integral, which is, of course, the area under the curve, is 
clearly divergent in the $r\rightarrow \infty$ limit.}
\label{fig7}
\end{figure}

\begin{figure}
\centering
\includegraphics[width=0.5\linewidth]{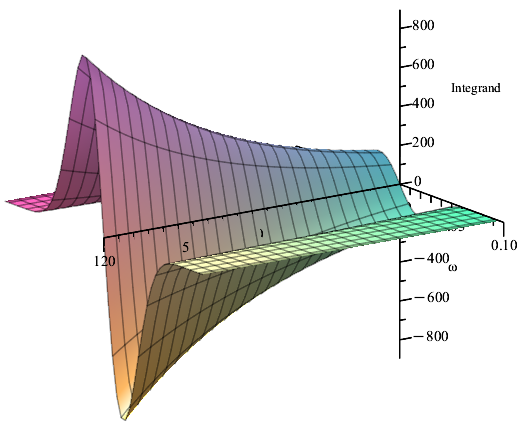}
\caption{{Plot}
 of Im$\left[\Omega_{0 \mu}^{II}(1,r,\omega)\right]$ for $r\in[100,120]$, 
$\omega \in [-\mu, \mu]$, setting $\mu=0.1$ and $M=10$, up~to the $r^{-5}$ term. 
Unlike the real part, the~phase shows a different behavior by virtue of 
its odd parity with respect to $\omega$. The~contributions to the integral 
on the $r$-axis seem to vanish in light of this feature.}
\label{fig8}
\end{figure}

The investigation of these two integrals is a hard task. 
However, we can rely on analytical approximations and numerical 
analysis in order to extract the essential features from them. The~
starting point is given by studying the behavior in frequencies 
of the two~integrand functions, which we label as

\begin{equation}
\Omega_{l \mu}^{I}(t,r,\omega)=\frac{1}{\sqrt{2\pi r(r-2M)}}
e^{-i \varphi(t,r,\omega)}\left(1+\sum_{s=1}^{\infty}
\frac{B_{s}(l,\omega,\mu)}{r^{s}}\right),
\label{(7.5)}
\end{equation}

\begin{equation}
\Omega_{l \mu}^{II}(t,r,\omega)=\frac{1}{\sqrt{2\pi r(r-2M)}}
e^{-i\omega t}e^{r\sqrt{\mu^2-\omega^2}
\left(1+M\frac{\mu^2-2\omega^2}{\sqrt{\mu^2-\omega^2}}
\frac{\log(r)}{r}\right)}\left(1+\sum_{s=1}^{\infty}
\frac{B_{s}(l,\omega,\mu)}{r^{s}}\right).
\label{(7.6)}
\end{equation}
{Notice} that the $l$-dependence is encoded in the coefficients of the 
series expansion, requiring to add also the angular part (given by 
spherical harmonics) to study the superposition of several waves. We therefore 
limit ourselves to the study of the $s$-wave, given by the constraint 
$B_s(l,\omega,\mu)=B_s(0,\omega,\mu)$. Moreover, for~numerical reasons, we 
consider only the first five negative powers of $r$, arriving at a 
simplified version of Equations \eqref{(7.5)} and \eqref{(7.6)}. Upon~setting
\begin{equation}
\mu=0.1, \quad t=1, \quad M=10,
\end{equation}
we first consider the plots of the real (Figure  \ref{fig7}) and imaginary 
(Figure  \ref{fig8}) part of Equation~\eqref{(7.6)}.
\vspace{-9pt}

It is possible to evaluate numerically the integral given by Equation 
\eqref{(7.4)} by expanding $ \Omega_{l \mu}^{II}(t,r,\omega)$ around $\omega =0$, 
since this region gives the largest contribution to it (as shown in
Figure~\ref{fig7}). We have performed such evaluation up to 
the sixth power of $\omega$ for the chosen parameters, 
arriving at the conclusion that such a piece 
diverges. In~order to obtain also an analytical estimate of such a divergence, 
we consider the change of variable given by
\begin{equation}
\omega=\mu\sin(\tau)
\end{equation}
applied to the leading term of Equation \eqref{(7.6)} and therefore obtain 
\begin{equation}
\Omega_{l \mu}^{II}(t,r,\tau)\approx\frac{1}{\sqrt{2\pi r(r-2M)}}r^{\mu M(2
\left|\cos(\tau)\right|-\text{sign}(\cos(\tau))\sec(\tau))}
e^{\mu(r\left|\cos(\tau)\right|-i\sin(\tau))}.
\end{equation}
{The} requirement $\omega \ll 1$ yields $\tau\ll1$, giving as leading term
\begin{equation}
\Omega_{l \mu}^{II}(t,r,\tau)\approx \frac{1}{\sqrt{2\pi r(r-2M)}} 
r^{\mu M}e^{\mu r}(i\mu t \tau -1) + \text{O}(\tau^2).
\end{equation}
{After} a straightforward integration we arrive at the desired result, i.e.,
\begin{equation}
\int_{-\frac{\pi}{2}}^ {\frac{\pi}{2}}d\tau \mu \cos(\tau) 
\Omega_{l \mu}^{II}(t,r,\tau)
\approx \frac{1}{\sqrt{2\pi r (r-2M)}}\mu e^{\mu r}r^{\mu M} +\text{O}(\tau^2).
\end{equation}
{Even} though it is possible to perform more accurate computations (including 
more terms and/or the negative powers of $r$ of the series expansion) it 
is clear that the dominant contribution is the one of the exponential term, 
which causes the divergence of the whole integral in the 
$r\rightarrow \infty$ limit, as~shown by the purely numerical analysis given 
before. Notice that the field mass term, $\mu$, plays a key role: in 
the massless case the equivalent of such a piece vanishes (as it happens here 
in the $\mu \rightarrow 0$ limit). We have to, therefore, face a case where a 
field not only does not vanish at spacelike infinity, but~furthermore it 
diverges exponentially. As~far as we can see, this property is compatible
with the findings in Ref.~\cite{Huh}, where part (II) of Proposition 2.1
therein shows that the norm of a massive scalar field obeying the
wave equation in Schwarzschild space-time 
is majorized by the product of two functions, one of
which is indeed divergent in a suitable limit. 
This agreement merits explicit mention because~the work in Ref.
~\cite{Huh} has exploited advanced methods from the modern theory
of hyperbolic equations, whereas we have used the standard techniques
of classical mathematical~physics.

\begin{figure}
\centering
\includegraphics[width=0.54\linewidth]{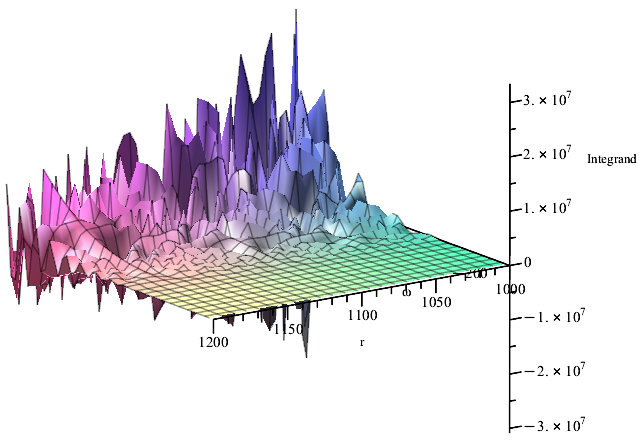}
\caption{{Plot} 
 of Re$\left[\Omega_{0 \mu}^{I}(1,r,\omega)\right]$ for 
$r\in[1, 1.2]\cdot10^3$, $\omega \in [-10^3, -\mu]$, setting $\mu=0.1$ and 
$M=10$, up~to the $r^{-5}$ term. For~every fixed value of $r$, the~field 
exhibits increasing oscillations as $\omega$ grows (in this specific case 
for $\omega<-400$). In~the low-frequency limit, the~field is negligible. 
Both behaviors suggest that the numerical interval is not divergent.}
\label{fig9}
\end{figure}

\begin{figure}
\centering
\includegraphics[width=0.52\linewidth]{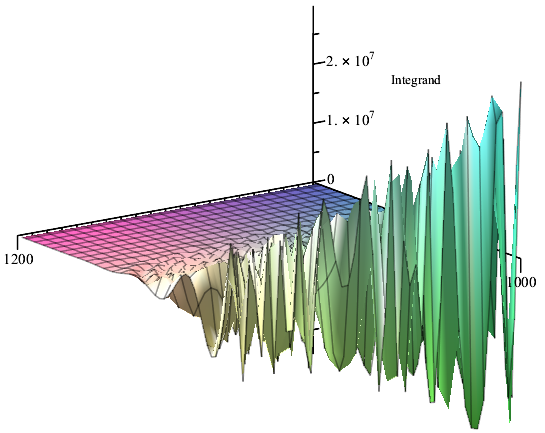}
\caption{Plot
 of Re$\left[\Omega_{0 \mu}^{I}(1,r,\omega)\right]$ for 
$r\in[1, 1.2]\cdot10^3$, $\omega \in [\mu, 10^3]$, setting $\mu=0.1$ and 
$M=10$, up~to the $r^{-5}$ term. The~discussion of such a figure follows 
closely the one of Figure~\ref{fig9}.}
\label{fig10}
\end{figure}

The case of $ \Omega_{l \mu}^I(t,r,\omega)$ is strongly different. 
Upon relying on the same conventions, we show the plots of its real part 
(the ones for the phase are very similar) for $\omega \in [-10^3, -\mu]$ in 
Figure~\ref{fig9} and for $\omega\in[\mu,10^3]$ in Figure~\ref{fig10}.

On both numerical and analytical grounds, we can say that, while the function 
$\Omega_l^I(t,r,\omega)$ diverges, the~associated integral (given by Equation \eqref{(7.3)}) 
remains finite and eventually vanishes at spacelike infinity. In~fact, we 
have an oscillating term $e^{-i\varphi}$ alleviated by a series 
of negative powers of $r$. 
  
Interestingly, we find that a massive scalar field contains a term that 
diverges at spacelike infinity, while this feature is absent in the massless case. 
In conclusion, the~mass plays a crucial role in the asymptotic description of 
a scalar field in spherical~coordinates.

\unskip

\section{Application of Section~\ref{666} to the Zerilli~Equation}\label{888}

The systematic {algorithm} of Section~\ref{666} can be applied to equations of even greater
interest. For~this purpose, we here consider the recent analysis in Ref.
~\cite{Hamaide} of the Zerilli equation~\cite{Zerilli}, which is obtained by perturbing a
background Schwarzschild metric with a small parity-even perturbation 
expressed by a symmetric tensor field of type $(0,2)$. Eventually, the~work
in Ref.~\cite{Hamaide} obtains an inhomogeneous radial wave equation, that
we express in the form
\begin{equation}
\left[\frac{d^{2}}{dr^{2}}+p(r)\frac{d}{dr}+q(r,l,\omega)\right]u(r,l,\omega)
={\widetilde I}(r,l,\omega),
\label{(8.1)}
\end{equation}
where
\begin{equation}
p(r)=\frac{2M}{r(r-2M)},
\label{(8.2)}
\end{equation}
\begin{equation}
q(r,l,\omega)=\frac{r^{2}}{(r-2M)^{2}}
(\omega^{2}-V(r,l,\omega)),
\label{(8.3)}
\end{equation}
having defined
\begin{equation}
V(r,l,\omega)=\left(1-\frac{2M}{r}\right)
\frac{[2\eta^{2}(\eta+1)r^{3}+6\eta^{2}Mr^{2}+18\eta M^{2}r
+18M^{3}]}{r^{3}(\eta r+3M)^{2}},
\label{(8.4)}
\end{equation}
with $\eta=\frac{(l-1)(l+2)}{2}$. The~work in Ref.~\cite{Hamaide} uses
a Green-function method in order to solve the inhomogeneous differential
equation \eqref{(8.1)} by using the well-established technique for obtaining
the Green function of the operator on the left-hand side of Equation \eqref{(8.1)} from solutions of the associated homogeneus equation
\begin{equation}
\left[\frac{d^{2}}{dr^{2}}+p(r)\frac{d}{dr}+q(r,l,\omega)\right]U(r,l,\omega)=0.
\label{(8.5)}
\end{equation}
{It} has been our choice to focus on this particular aspect of the general
problem since we have already written several sections on the massive
scalar case, which is the main topic of our~paper. 

By repeating the reduction to normal form as we did in {Section}~\ref{222}, we obtain
$U(r,l,\omega)$ in the factorized form
\begin{equation}
U(r,l,\omega)=\gamma(r)\beta(r,l,\omega),
\label{(8.6)}
\end{equation}
where
\begin{equation}
\gamma(r)=\frac{1}{\sqrt{1-\frac{2M}{r}}},
\label{(8.7)}
\end{equation}
while $\beta(r,l,\omega)$ solves the equation
\begin{equation}
\left[\frac{d^{2}}{dr^{2}}+J(r,l,\omega)\right]\beta(r,l,\omega)=0,
\label{(8.8)}
\end{equation}
having defined the potential term
\begin{eqnarray}
\; & \; & J(r,l,\omega)=\frac{1}{(r-2M)^{2}r^{2}[(l(l+1)-2)r+6M]^{2}}
\Bigr[\omega^{2}(l-1)^{2}(l+2)^{2}r^{6}
\nonumber \\
&+& 12M \omega^{2}(l-1)(l+2)r^{5}
+(-l^{6}-3l^{5}+l^{4}+7l^{3}-4l+36M^{2}\omega^{2})r^{4}
\nonumber \\
&+& 2M(l-1)^{3}(l+2)^{3}r^{3}
+9M^{2}\left(l(l+1)-\frac{10}{3}\right)(l+2)(l-1)r^{2}
\nonumber \\
&+& 36M^{3}(l-1)(l+2)r+36M^{4}\Bigr].
\label{(8.9)}
\end{eqnarray}
{At} this stage, in~close analogy with Equation \eqref{(6.1)}, we may expand
Equation \eqref{(8.8)} as $r \rightarrow \infty$ in the form
\begin{equation}
\left(\frac{d^{2}}{dr^{2}}+\sum_{k=0}^{\infty}
\frac{A_{k}(l,\omega)}{r^{k}}\right)\beta(r,l,\omega)=0,
\label{(8.10)}
\end{equation}
where the function $\beta$ has the asymptotic expansion
\begin{equation}
\beta(r,l,\omega) \sim e^{- r \sqrt{-A_{0}}} \; r^{\zeta}
\left(1+\sum_{s=1}^{\infty}\frac{B_{s}(l,\omega)}{r^{s}}\right).
\label{(8.11)}
\end{equation}
{The} formulae for expressing $\zeta$ and $B_{s}$ in terms of the
$A_{k}(l,\omega)$ coincide with the ones in Section~\ref{666}, whereas the
$A_{k}$ are different. Indeed, we find for example that
\begin{equation}
A_{0}=\omega^{2},
\label{(8.12)}
\end{equation}
\begin{equation}
A_{1}=4 \omega^{2}M,
\label{(8.13)}
\end{equation}
\begin{equation}
A_{2}=12M^{2}\omega^{2}-l(l+1),
\label{(8.14)}
\end{equation}
\begin{equation}
A_{3}=\frac{2M[16M^{2}\omega^{2}(l(l+1)-2)-l^{4}-2l^{3}
+5l^{2}+6l+4]}{l(l+1)-2},
\label{(8.15)}
\end{equation}
and hence, from~Equations \eqref{(6.7)} and \eqref{(6.8)},
\begin{equation}
\zeta=-2iM |\omega|,
\label{(8.16)}
\end{equation}
\begin{equation}
B_1=-M + \frac{i[8M^2 \omega^2-l(l+1)]}{2|\omega|},
\label{(8.17)}
\end{equation}
supplemented by a countable infinity of higher-order 
$A_{k}$ and $B_s$ coefficients. 

\begin{figure}
\centering
\includegraphics[width=0.54\linewidth]{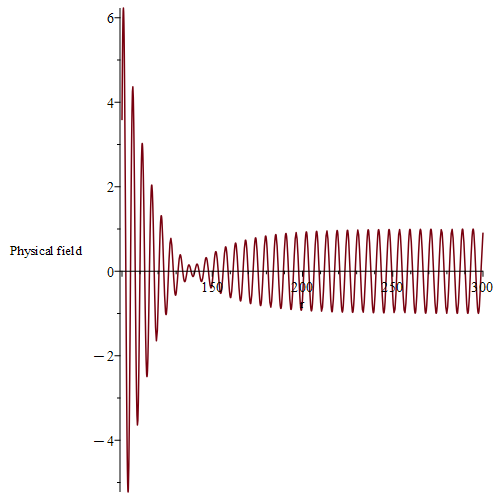}
\caption{The real part of Equation \eqref{(8.6)} is shown for $r\in [100,300]$, 
given the parameters \eqref{(8.18)} and summing over $s$ from 1 to 5
in Equation \eqref{(8.11)}. 
After achieving high values in the region closest to the event horizon, 
the field shrinks at $r\approx 140$ and then behaves like a plane wave, as~it should be.}
\label{fig11}
\end{figure}

\begin{figure}
\centering
\includegraphics[width=0.53\linewidth]{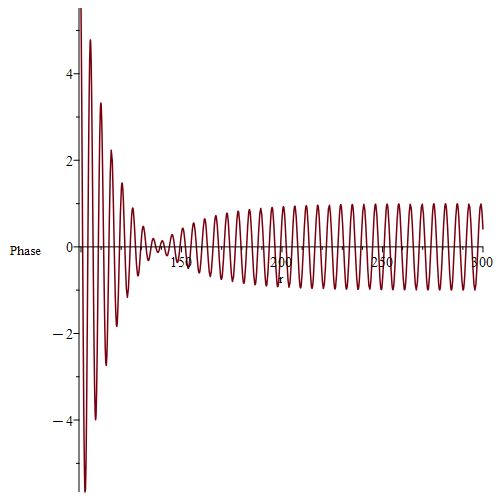}
\caption{The imaginary part of Equation \eqref{(8.6)} is shown for $r\in [100,300]$, 
given the parameters \eqref{(8.18)} and summing over $s$ from 1 to 5
in Equation \eqref{(8.11)}. Its 
behavior is very similar to the one of the real part, leading to the same conclusions.}
\label{fig12}
\end{figure}

Hereafter, we plot the real (Figure  \ref{fig11}) and imaginary (Figure  \ref{fig12}) 
part of Equation \eqref{(8.6)} upon considering the parameters
\begin{equation}
M=10, \quad \omega=1, \quad l=0,
\label{(8.18)}
\end{equation}
up to the $B_5$ coefficient in Equation \eqref{(8.11)}. 
\unskip

Figures~\ref{fig11} and \ref{fig12} show in a simple way that our findings 
are in full agreement with the plane-wave
behavior as $r \rightarrow \infty$ obtained in Ref.~\cite{Hamaide} since,
after the shrinking of the field at $r \simeq 140$, both
real and imaginary part of the solution display constant oscillations through spacelike~infinity.

\section{Results Obtained and Open~Problems}\label{999}

The scalar wave equation in Schwarzschild space-time has been studied
intensively over more than half a century~\cite{Persides1,Persides2,Persides3,Schmidt,Fiziev,Philipp,Kehrberger},
until the very recent and highly valuable work in Ref.
~\cite{Bianchi}, but~the material in our Sections~\ref{666} and~\ref{777} remains original.
More precisely, as~far as we know, our original contributions 
deal with ordinary and partial differential equations 
for black hole space-times and are as {follows:}

\begin{enumerate}
\item [(i)] For a massive scalar field in Schwarzschild geometry, our Equations 
\eqref{(6.7)}--\eqref{(6.9)} and \eqref{(C12)}--\eqref{(C13)}
yield the full Poincar\'e asymptotic expansion of Fourier modes that
solve the radial wave~equation.

\item [(ii)] Our Equations \eqref{(7.3)}--\eqref{(7.4)} yield the integral representation
of the massive scalar field, with~the resulting numerical analysis
discussed therein.
\end{enumerate}

Our formulae provide useful insights into the physical behavior of a massive 
scalar field in Schwarzschild space-time as the singular points of the
radial wave equation are approached. Recognizing the radial equation as
nothing but a particular case of the confluent Heun equation makes it
possible to rely upon established mathematical techniques without 
approximations. The~{algorithm} developed in Sections~\ref{666} and~\ref{777} allows,
in principle, to~compute every term of the series at infinity by reaching
the desired level of accuracy. Moreover, even if we have shown plots 
only for specific values of parameters, our formulae can be easily implemented
in every symbolic programming~language.

For massive scalar fields in Schwarzschild space-time, the~literature
has mainly focused on late-time behavior~\cite{Burko} of Fourier modes, whereas rigorous results 
on solutions of the Cauchy problem are along the lines of Ref.~\cite{Huh}.
We think that the next step will consist in applying analogous techniques to 
other spherically symmetric space-times, both in black-hole theory~\cite{Minucci} and in cosmological backgrounds~\cite{BE}.
For this purpose, we have eventually studied the homogeneous equation 
associated with the inhomogeneous Zerilli equation in a Schwarzschild
background. Our original Formulas 
\eqref{(8.12)}--\eqref{(8.17)} can be applied to build the
Green function, which in turn leads to the solution of the inhomogeneous
Zerilli equation as shown in Ref.~\cite{Hamaide}.

More precisely, the~following remarks are still in {order:} 

\begin{enumerate}
\item [(i)] The exponential divergence of the massive field at spacelike 
infinity arises from the integral in Equation \eqref{(7.4)}.
As far as we can see, no boundary condition can get rid of it. As~
we stressed in {Section}~\ref{777}, it is encouraging that our result agrees
with a more advanced mathematical analysis as the one performed
in Ref.~\cite{Huh}, which had been ignored in the physics-oriented
literature. Maybe this property means that one has instead to
investigate the coupled system consisting of Einstein and Klein--Gordon
equations, but~this would require a separate paper.
\item [(ii)] The proof of convergence or divergence of the series in Equation \eqref{(6.9)} is an open and difficult technical problem
because the $B_s$ coefficients therein contain rational functions
of the parameters and products of such rational functions with
square roots of polynomials in the parameters. However,
as we say in Appendix \ref{appE}, an~even more fundamental open problem
is whether one should keep using the Poincar\'e definition of
asymptotic expansion. This topic deserves a separate paper as well.
Anyway, for~example, the~Poincar\'e asymptotic expansion of the logarithm of the
$\Gamma$-function is reliable at large values of the independent
variable but is expressed by a divergent 
series~\cite{Poincare2}. Thus, good 
asymptotic estimates at large $r$ in our paper would not be affected
by divergence of the expansion.
\item [(iii)] The cases $\omega^{2} < \mu^{2}$ and $\omega^{2} > \mu^{2}$
studied in {Section}~\ref{777} are both physically relevant, and~we cannot
foresee a viable regularization for the time being. Maybe one has
to revert to what we have suggested at the end of item (i) in the
present list.
\item [(iv)]The familiarity with evaluation of $B_s$ coefficients 
gained in {Section}~\ref{666} will actually make it easier to complete the
analysis of {Section}~\ref{888}. More precisely, the~$B_s$ coefficients
in Equation \eqref{(8.11)} do not depend on the mass of any field 
and are simpler in this respect, but~on the other 
hand, they have both real and imaginary
parts, as~is clear from Equation~\eqref{(8.17)}. Their evaluation
by computer programs will lead to the evaluation of Fourier modes with 
the desired accuracy, and~in turn, the Green function of the
inhomogeneous radial wave equation \eqref{(8.1)} will be obtained
with the same accuracy.
\item [(v)] We cannot foresee any obstruction to studying wave equations
in a Kerr background with our technique. The~modern packages
will help a lot in evaluating $B_s$ coefficients which depend
also on angular momentum. Once more, we conclude that a separate
paper is necessary to accomplish this task as well.
\end{enumerate}
\vspace{6pt}

{{The}
authors have equally contributed to conceptualization, methodology, validation, formal analysis.
G.E. was the expert in draft preparation, while M.R. was the software expert.
All authors have read
and agreed to the published version of the~manuscript.}

{This research received no external~funding.}

{{No} new data were created.}

\acknowledgments{G.E. is grateful to INDAM for membership. 
G.E. dedicates the paper to his daughter Margherita, and~M.R. dedicates this work to his mother~Tania.}
\\
{The authors declare no conflict of~interest.}
\appendix
\section[\appendixname~\thesection]{The Confluent Heun Equation and Its Solutions}\label{appA}

Given a differentiable function $Y$ of the variable $z$, the~confluent
Heun equation~\cite{Ronveaux} is a second-order ordinary differential equation 
of the following form:
\begin{equation}
\left[\frac{d^{2}}{dz^{2}}
+\left(\alpha+\frac{(\beta+1)}{z}
+\frac{(\gamma+1)}{(z-1)}\right)\frac{d}{dz} 
+ \left(\frac{\mu}{z}
+\frac{\nu}{(z-1)}\right) \right]Y(z)=0,
\label{(A1)}
\end{equation}
where the parameters $\mu$ and $\nu$ obey the relations
\begin{equation}
\mu=\frac{1}{2}(\alpha-\beta-\gamma-2 \eta+\alpha \beta-\beta \gamma),
\label{(A2)}
\end{equation}
\begin{equation}
\nu=\frac{1}{2}(\alpha+\beta+\gamma+2 \delta + 2 \eta
+\alpha \gamma + \beta \gamma),
\label{(A3)}
\end{equation}
while $\alpha,\beta,\gamma,\delta,\eta$ are free (in general complex)
parameters. A~Heun equation has four singular points, while its 
confluent form has two regular singular points at $z=0$, $z=1$ 
of $s$-rank $1$ and an irregular singular point at $z=\infty$ of
$s$-rank $2$. Indeed, the~confluent form is obtained from the standard
one by gluing two singularities. In~the literature, only power-series
solutions in the neighborhood of singular points are known, resulting
in three sets of two linearly independent solutions. The~ones in the
neighborhood of $z=0$ and $z=1$ are called Frobenius solutions, while
the ones in the neighborhood of the point at infinity are known as
Thom\'e solutions. Upon~denoting with
$$
{\rm HeunC}(\alpha,\beta,\gamma,\delta,\eta,z)
$$
the first Frobenius solution as $z$ approaches $0$, we have that
\begin{equation}
Y_{0}(z)=C_{1}{\rm HeunC}(\alpha,\beta,\gamma,\delta,\eta,z)
+ C_{2}z^{-\beta}{\rm HeunC}(\alpha,-\beta,\gamma,\delta,\eta,z)
\label{(A4)}
\end{equation}
is the complete Frobenius solution as $z \rightarrow 0$, while
\begin{eqnarray}
\; & \; & Y_{1}(z)=C_{1}
{\rm HeunC}(-\alpha,\gamma,\beta,-\delta,\eta+\delta,1-z) 
\nonumber \\
&+& C_{2} (z-1)^{-c}{\rm HeunC}
(-\alpha,-\gamma,\beta,-\delta,\eta+\delta,1-z)
\label{(A5)}
\end{eqnarray}
is the complete Frobenius solution as $z \rightarrow 1$. Their leading
order is given by
\begin{equation}
Y_{0}(z) \sim C_{1}+C_{2}z^{-\beta},
\label{(A6)}
\end{equation}
\begin{equation}
Y_{1}(z) \sim C_{1}+C_{2}(z-1)^{-c}.
\label{(A7)}
\end{equation}
{The} function ${\rm HeunC}$ is implemented in many symbolic computer programs
such as Mathematica and Maple. In~our paper we are adopting Maple
conventions. Once that ${\rm HeunC}$ is known, it is straightforward to
build the two sets of linearly independent Frobenius solutions written 
before. Their radius of convergence is obtained from the conditions that
the absolute value of the argument of ${\rm HeunC}$ must be less
than $1$: $|z| <1$. The~solution for all values of $z$ is obtained by
overlapping the analytical continuations of $Y_{0}(z)$ and $Y_{1}(z)$.

The solution at infinity is quite different. It is well known that every 
Heun equation (also the confluent, biconfluent and three-confluent one)
admits a solution at infinity whose asymptotic expansion is given by
\begin{equation}
Y_{\infty}(z)=e^{T_{n}(z)} \; z^{-\theta} \;
\sum_{k=0}^{\infty}\rho_{k}z^{-k},
\label{(A8)}
\end{equation}
where $T_{n}$ is a polynomial function of $z$. Note that the sign of
$\theta$ may change depending on the chosen parametrization. Upon~setting $T_{n}(z) \sim 1+z$ one obtains the so-called Thom\'e~solutions
\begin{equation}
Y_{\infty}(z)= C_{1}z^{-\theta_{1}} \sum_{k=0}^{\infty} \rho_{k}z^{-k}
+ C_{2}e^{\theta_{2}z} \; z^{-\theta_{3}} \;
\sum_{k=0}^{\infty}\sigma_{k}z^{-k}.
\label{(A9)}
\end{equation}
{By} inserting them into Equation \eqref{(A1)} it is possible to obtain a
recurrence relation for $\rho_{k},\sigma_{k}$. Note that 
$\theta_{1},\theta_{2},\theta_{3}$ are in principle different and
complex. However, it is possible to relate them to 
$\alpha,\beta,\gamma,\delta,\eta$ even though this strongly depends
on the chosen~parametrization.

\section[\appendixname~\thesection]{Canonical Reduction with Respect to \boldmath{$z$}}\label{appB}

It is very easy to perform the canonical reduction of the radial equation
\eqref{(2.9)} with respect to the independent variable $z$. Following the
same procedure of {Section}~\ref{222}, we obtain
\begin{equation}
\alpha(z)=\frac{1}{\sqrt{z(1-z)}},
\label{(B1)}
\end{equation}
while the potential term reads as
\begin{eqnarray}
\; & \; & T_{l \omega \mu}(z)
=\frac{(1+16M^{2}\omega^{2})}{4z^{2}(z-1)^{2}}
+\frac{(4M^{2}(\mu^{2}-4\omega^{2})+l(l+1))}{z(z-1)^{2}}
\nonumber \\
&+& \frac{(12M^{2}(2\omega^{2}-\mu^{2})-l(l+1))}{(z-1)^{2}}
+\frac{4M^{2}(3\mu^{2}-4 \omega^{2})z}{(z-1)^{2}}
\nonumber \\
&+& \frac{4M^{2}(\omega^{2}-\mu^{2})z^{2}}{(z-1)^{2}}.
\label{(B2)}
\end{eqnarray}
{The} function $\beta_{l \omega \mu}$ is the solution of the differential equation
\begin{equation}
\left[\frac{d^{2}}{dz^{2}}+T_{l \omega \mu}(z)\right]\beta_{l \omega \mu}(z)=0.
\label{(B3)}
\end{equation}
{In} the neighborhood of $z=0$, the~potential takes the approximate form
\begin{equation}
T_{0 l \omega}(z) \sim \left(4M^{2}\omega^{2}+\frac{1}{4}\right)z^{-2},
\label{(B4)}
\end{equation}
which leads to the approximate solution
\begin{equation}
\beta_{0 l \omega}(z)=D_{1}z^{2i M \omega + \frac{1}{2}}
+D_{2}z^{-2iM \omega + \frac{1}{2}}.
\label{(B5)}
\end{equation}
{In} the neighborhood of $z=1$ we find instead (after imposing
regularity of $\beta$ therein)
\begin{equation}
T_{1 l \omega}(z) \sim \frac{1}{4} (z-1)^{-2},
\label{(B6)}
\end{equation}
\begin{equation}
\beta_{1 l \omega}(z)=C_{1}\sqrt{z-1}.
\label{(B7)}
\end{equation}
{At} infinity (i.e., as~$r$ approaches $+\infty$, which implies that
$z$ approaches $-\infty$) we find
\begin{equation}
T_{\infty l \omega \mu}(z) \sim 4M^{2}(\omega^{2}-\mu^{2}),
\label{(B8)}
\end{equation}
\begin{equation}
\beta_{\infty l \omega \mu}(z)=S_{1}e^{2zM \sqrt{\omega^{2}-\mu^{2}}}
+S_{2}e^{-2z \sqrt{\omega^{2}-\mu^{2}}}.
\label{(B9)}
\end{equation}
{We} note complete correspondence with the leading behaviors obtained
in {Section}~\ref{333} with respect to the $r$ variable. 

\section[\appendixname~\thesection]{Remarks on the Solution at Infinity}\label{appC}

First, we provide a theoretical derivation of the ansatz given by Equation 
\eqref{(6.2)} based on a method developed by Poincar\'e~\cite{Poincare,Poincare2}
to compute the solution of a differential equation with variable coefficients. 
We then evaluate the explicit form of the first coefficients 
$B_k$ of the solution given by Equation \eqref{(6.9)}.

\subsection[\appendixname~\thesection]{Theoretical Derivation}

The radial wave equation in its normal form \eqref{(2.9)}, with~the potential
term defined in Equation \eqref{(2.10)}, can be solved locally, i.e.,~in the
neighborhood of the singular\linebreak   points~\cite{Persides1,Fiziev,Philipp,Minucci}. 
For this purpose,
it is helpful to re-express it in a form where the variable coefficients
are polynomials, following the seminal work of Poincar\'e~\cite{Poincare}.
This is achieved by multiplying it on the left by $r^{2}(r-2M)^{2}$. 
This leads to the differential equation
\begin{equation}
\left[P_{2}(r)\frac{d^{2}}{dr^{2}}
+P_{0l\omega \mu}(r)\right]\beta_{l\omega \mu}(r)=0,
\label{canonical Poincare 1}
\end{equation}
where
\begin{equation}
P_{2}(r)=r^{2}(r-2M)^{2}=r^{4}-4Mr^{3}+4M^{2}r^{2},
\label{canonical Poincare 2}
\end{equation}
\begin{equation}
\begin{array}{lll}
\; & \; & P_{0l\omega \mu}(r)=r^{2}(r-2M)^{2}J_{l\omega \mu}(r)
 \\
&=& (\omega^{2}-\mu^{2})r^{4}+2M\mu^{2}r^{3}
+\left[\frac{1}{2}-\frac{1}{2}-l(l+1)\right]r^{2}
 \\
&+& \Bigr[-M+M+2Ml(l+1)\Bigr]r+M^{2}
 \\
&=& (\omega^{2}-\mu^{2})r^{4}+2M \mu^{2}r^{3}
-l(l+1)r^{2}+2Ml(l+1)r+M^{2},
\label{canonical Poincare 3}
\end{array}
\end{equation}
and we have stressed in the intermediate step the exact cancellations 
occurring in the evaluation of the polynomial 
\eqref{canonical Poincare 3}. As~far as we know, only the authors of Ref.~\cite{Philipp} came pretty close to studying an equation like our
\eqref{canonical Poincare 1} with $r$ as an independent variable, 
but they chose a third degree polynomial in their Equation~(32) as the 
coefficient of the second derivative of the desired~solution.

At this stage, the~approximate solutions \eqref{(3.3)}, \eqref{(3.13)} 
and the one which would be obtained for Equation \eqref{(3.14)} by considering only 
the leading term (labelled as $\beta_{\omega,\infty}$) 
suggest looking for a solution of Equation 
\eqref{canonical Poincare 1} in the factorized form of product (called
$s$-homotopic ansatz in the current literature~\cite{Philipp,Minucci})
\begin{equation}
\beta_{l\omega \mu}(r)=\beta_{0}(r)\beta_{\omega,2M}(r)
\beta_{\omega,\infty}(r)G_{l\omega \mu}(r),
\label{canonical Poincare 4}
\end{equation}
where the function $G_{l\omega \mu}$ should be such that this product solves,
in the neighborhood of $r=0,2M,\infty$, Equation \eqref{canonical Poincare 1}. In~the ansatz 
\eqref{canonical Poincare 4} we have
\vspace{-6pt}
\begingroup\makeatletter\def\f@size{9}\check@mathfonts
\def\maketag@@@#1{\hbox{\m@th\normalsize\normalfont#1}}
\begin{equation}
\begin{array}{lll}
\; & \; & \beta_{0}(r)\beta_{\omega,2M}(r)\beta_{\omega,\infty}(r)
=\sqrt{r(r-2M)} \Bigr[D_{1}(r-2M)^{2iM \omega}
+D_{2}(r-2M)^{-2iM\omega}\Bigr]
 \\
& \times & \Bigr[S_{1}e^{r \sqrt{\mu^{2}-\omega^{2}}}
+S_{2}e^{-r \sqrt{\mu^{2}-\omega^{2}}}\Bigr].
\label{canonical Poincare 5}
\end{array}
\end{equation}
\endgroup
{The} factorization in Equation \eqref{canonical Poincare 4} follows what was learned from
Poincar\'e~\cite{Poincare2}, i.e.,~for linear differential equations
with variable coefficients, the~solution consists of the limiting
form at the origin times the limiting form at infinity times a suitable
series. With~respect to the case considered in Ref.~\cite{Poincare2},
we are here dealing with two (rather than just one) regular singular 
points at finite values of the independent variable. The~structure of the 
ansatz given by Equations \eqref{canonical Poincare 4}, \eqref{canonical Poincare 5} 
is too rich, and it does not provide any simplification of computations. 
Our proposal, given by Equation \eqref{(6.2)}, is an easier case where 
$S_1=0$, $\beta_{0}(r)\beta_{\omega,2M}(r)\simeq r^\alpha$ and 
the interpolating series $G_{l\omega \mu}(r)$ consists only of negative powers of r. 
In this way we are able to study the behavior of the field at infinity without any~effort.

\subsection[\appendixname~\thesection]{Some Coefficients}

We begin by writing the first coefficients $A_k$ of the asymptotic expansion 
of the canonical potential at infinity, given by Equation \eqref{(3.14)}:
\begin{equation}
A_0=\omega^2-\mu^2,
\end{equation}
\begin{equation}
A_1=2M(2 \omega^{2}-\mu^{2}),
\end{equation}
\begin{equation}
A_2=4M^{2}(3 \omega^{2}-\mu^{2})-l(l+1),
\end{equation}
\begin{equation}
A_3=2M[16\omega^2M^2-4\mu^2M^2-l(l+1)],
\end{equation}
\begin{equation}
A_4=M^2[80\omega^2M^2-16\mu^2M^2-4l(l+1)+1],
\end{equation}
\begin{equation}
A_5=4M^3[48M^2\omega^2-8\mu^2M^2-2l(l+1)+1].
\end{equation}

In light of the above, the~first two $B_s$ coefficients are explicitly 
obtained here:
\begin{equation}
\begin{split}
B_1=\frac{1}{2(\mu^2-\omega^2)^2}& \biggr \{\sqrt{\mu^2-\omega^2}\Bigr[3M^2\mu^4
+\mu^2(l(l+1)-12\omega^2M^2)+8M^2\omega^4 \\ 
&-l(l+1)\omega^{2}\Bigr]
-M\mu^4+3M\mu^2\omega^2-2M\omega^4 \biggr \},
\end{split}
\label{(C12)}
\end{equation}
\begin{equation}
\begin{split}
B_2=&\frac{1}{2(\mu^2-\omega^2)^3} \biggr \{4M\sqrt{\mu^2-\omega^2}
\Bigr[\omega^4(4M^2\mu^2-l(l+1)+1) \\
&-6\omega^2\mu^2\left(M^2\mu^2-\frac{1}{6}l(l+1)
+\frac{1}{4}\right)+M^2\mu^6+\frac{\mu^4}{2}\Bigr]\\
&+64M^4\omega^8
-192M^2\omega^6\left(M^2\mu^2+\frac{1}{12}l(l+1)-\frac{1}{48}\right)\\
&+\omega^4 \Bigr[192M^4\mu^4+40M^2\mu^2\left(l(l+1)-\frac{2}{5}\right)+l^4+2l^3-l^2-2l \Bigr]\\
&-72\mu^2\omega^2 \Bigr[M^4\mu^4+\frac{5}{12}M^2\mu^2\left(l(l+1)-\frac{1}{2}\right)
+\frac{1}{36}l(l+2)(l^2-1)\Bigr]\\
&+9\mu^4\left(M^2\mu^2+\frac{1}{3}l^2
-\frac{1}{3}\right)\left(M^2\mu^2+\frac{l^2}{3}+\frac{2}{3}l\right) \biggr \}.
\end{split}
\label{(C13)}
\end{equation}
{For} $s>2$, the~coefficients start getting cumbersome. However, their evaluation remains
feasible by virtue of the {algorithm} described in the main~text.

\section[\appendixname~\thesection]{Comparison with the Persides Analysis}\label{appD}

{Sections \ref{666} and \ref{777} have presented our original calculations for a massive
scalar field in a fixed Schwarzschild background, whereas, for~a massless field, 
the large-$r$ solution of
the radial wave equation was first obtained by Persides~\cite{Persides1} by using a 
$3$-term recurrence relation, because~he did not reduce the equation
for Fourier modes $R_{l \omega}$ to its normal form. 
With our notation, the~Persides large-$r$ form of
$R_{l \omega}(r)$ reads as (setting $c$ and $G$ to $1$ in Ref.~\cite{Persides1})}
\begin{equation}
R_{l \omega,\pm}(r) \sim e^{\mp i \omega(r+2M\log(r-2M))}F_{l\omega,\pm}(r),
\label{(D1)}
\end{equation}
the function $F_{l \omega,\pm}$ being obtainable as
\begin{equation}
F_{l \omega,\pm}(r)=F_{l}(\omega r; \epsilon=\pm 2i),
\label{(D2)}
\end{equation}
where $F_{l}(\omega r; \epsilon)$ solves, by~definition, the~differential
equation (here $z=\omega r$)
\begin{equation}
\biggr[z(z-2M\omega)\frac{d^{2}}{dz^{2}}
+(-\epsilon z^{2}+2z-2M \omega)\frac{d}{dz}
-(\epsilon z +l(l+1))\biggr]F_{l}=0,
\label{(D3)}
\end{equation}
and has the large-$z$ (i.e., large-$r$) asymptotic expansion
\begin{equation}
F_{l}(\omega r; \epsilon) \sim \left(\frac{\epsilon}{2}\right)^{l+1}
\sum_{n=0}^{\infty}\tau_{n}(l,\omega)z^{-(n+1)},
\label{(D4)}
\end{equation}
where $\tau_{0}=1$, while~\cite{Persides1}
\begin{equation}
n \epsilon \tau_{n}-(l+n)(l-n+1)\tau_{n-1}
-2M\omega (n-1)^{2}\tau_{n-2}=0.
\label{(D5)}
\end{equation}
{The} explicit form of the first four $\tau_n$ obtained 
by such a relation in the general $\tau_0\neq1$ case is as follows:
\begin{equation*}
\begin{split}
\tau_1=&\frac{\tau_0}{\epsilon}(l+1)l,\\
\tau_2=&\frac{\tau_0}{2\epsilon^2}[(l+2)(l^2-1)l+ 2M\omega\epsilon ],\\
\tau_3=&\frac{\tau_0}{6\epsilon^3}[(l+3)(l^5-5l^3+4l)+6M\omega\epsilon(3l^2+3l-2)],\\
\tau_4=&\frac{\tau_0}{24\epsilon^4}[(l+4)(l^7-14l^5+49l^3-36l)
+12M\omega\epsilon(6l^4+9M\omega\epsilon+12l^3-22l^2-28l+12)].
\end{split}
\end{equation*}

On the other hand, in~light of our {Section}~\ref{666}, one might be tempted to
study the massless case by defining
$$
R_{l \omega,+}(r) \sim \frac{1}{\sqrt{r(r-2M)}}
e^{-i\omega(r+2M\log(r))}
\left(1+\sum_{s=1}^{\infty}\frac{B_{s}(l,\omega,0)}{r^{s}}\right).
$$
{However}, such a formula cannot be taken to hold, because~Equation \eqref{(6.9)} is only a Poincar\'e asymptotic expansion
(see Appendix \ref{appE}~\cite{Ronveaux}), and~hence
$$
\lim_{\mu \to 0} \sum_{s=1}^{\infty}
\frac{B_{s}(l,\omega,\mu)}{r^{s}}
\not =\sum_{s=1}^{\infty}\lim_{\mu \to 0}
\frac{B_{s}(l,\omega,\mu)}{r^{s}}.
$$
{As} far as we can see, the~best that one can do is to check that by~including as many $B_{s}(l,\omega,\mu)$ coefficients as possible
in the evaluation of the massive case, and~as many $\tau_{n}(l,\omega)$
coefficients as possible in the massless case, one obtains an accurate
approximation of the local solution of the radial wave equation in
the neighborhood of $r=\infty$. In~the latter case, one should also
bear in mind that~\cite{Persides1}
\begin{equation}
F_{l \omega,+}(r)=F_{l}(\omega r;\epsilon=2i) 
\sim i^{l+1}\sum_{n=0}^{\infty}\tau_{n}
(\omega r)^{-(n+1)}.
\label{(D6)}
\end{equation}

\section[\appendixname~\thesection]{The Poincar\'e Framework}\label{appE}

In Ref.~\cite{Poincare2}, Poincar\'e studied linear differential
equations of the form
\begin{equation}
\sum_{k=0}^{n}P_{k}(x)\frac{d^{k}y}{dx^{k}}=0,
\label{(E1)}
\end{equation}
where the $P_{k}(x)$ are polynomials in the variable $x$. He knew
from the work of Fuchs and Thom\'e that, apart from some exceptional
cases, there exist $n$ functions of the following~form:
\begin{equation}
e^{Q(x)} \; x^{a} \; \left(A_{0}+\frac{A_{1}}{x}
+\frac{A_{2}}{x^{2}}+...\right),
\label{(E2)}
\end{equation}
which satisfy formally Equation \eqref{(E1)}, where $Q(x)$ is a polynomial
in $x$. The~series in Equation \eqref{(E2)} was then said to be a normal 
series. In~particular, if~the normal series is convergent, one says 
that Equation \eqref{(E1)} admits a normal~integral. 

Poincar\'e says that a divergent series 
$$
\sum_{k=0}^{\infty}\frac{A_{k}}{x^{k}},
$$
where the sum of the first $(n+1)$ terms is denoted by $S_{n}$,
represents asymptotically a function $J: x \rightarrow J(x)$ if
the expression 
$$
x^{n}(J(x)-S_{n}(x))
$$
approaches $0$ when $x$ tends to infinity:
\begin{equation}
\lim_{x \to \infty}x^{n}(J(x)-S_{n}(x))=0.
\label{(E3)}
\end{equation}
{This} means that, for~$x$ sufficiently large, one has the majorization
\begin{equation}
x^{n}(J(x)-S_{n}(x)) < \varepsilon.
\label{(E4)}
\end{equation}
{The} error
\begin{equation}
J(x)-S_{n}(x)=\frac{\varepsilon}{x^{n}}
\label{(E5)}
\end{equation}
made on taking (just) the first $(n+1)$ terms is then extremely 
small, and~also much smaller than the error made on taking just 
$n$ terms, which equals
\begin{equation}
J(x)-S_{n-1}(x)=\frac{A_{n}+\varepsilon}{x^{n}},
\label{(E6)}
\end{equation}
$\varepsilon$ being very small and $A_{n}$ being~finite.

{The literature on theoretical physics still relies upon the
Poincar\'e definition of asymptotic expansion, whereas in pure
mathematics one has also a different, rigorous framework 
built by Dieudonn\'e~\cite{Dieudonne}.}

\end{document}